\documentclass[journal=apchd5,manuscript=article]{achemso}

\usepackage[version=3]{mhchem} 
\usepackage{layouts}
\author{Matthias Kraft}
\altaffiliation{These authors contributed equally}
\email{matthias.kraft09@imperial.ac.uk}
\affiliation[Imperial College London]{Blackett Laboratory, Department of Physics, Imperial College London, London SW7 2AZ, United Kingdom}
\author{Avi Braun}
\email{a.braun@imperial.ac.uk}
\altaffiliation{These authors contributed equally}
\affiliation[Imperial College London]{Blackett Laboratory, Department of Physics, Imperial College London, London SW7 2AZ, United Kingdom}
\author{Yu Luo}
\affiliation[Nanyang Technological University]{School of Electrical and Electronic
Engineering, Nanyang Technological University, Nanyang Avenue 639798, Singapore}
\author{S. A. Maier}
\affiliation[Imperial College London]{Blackett Laboratory, Department of Physics, Imperial College London, London SW7 2AZ, United Kingdom}
\author{J. B. Pendry}
\affiliation[Imperial College London]{Blackett Laboratory, Department of Physics, Imperial College London, London SW7 2AZ, United Kingdom}

\title{Bianisotropy and magnetism in plasmonic gratings}


\begin{document}
\begin{abstract}
We present a simple design to achieve bianisotropy at visible wavelengths: an
ultrathin plasmonic grating made of a gold grating covered by a thin flat layer
of gold. We show experimentally and through simulations that the grating
exhibits magneto-electric coupling and features asymmetric reflection and
absorption, all that with a device thickness of a tenth of the operating
wavelength. We compared the spectral results and retrieved the effective
material parameters of different polarizations and devices. We show that both
asymmetry and strong coupling between the incoming light and the optically
interacting surfaces are required for obtaining asymmetric optical behavior in
metasurfaces.
\end{abstract}

\section{Introduction}

In their seminal book \cite{Landau}, Landau and Lifshitz commented on the 
insignificance of magnetism at optical frequencies. Yet, several recent studies have
shown that magnetic effects are present in optical systems \cite{Shalaev2007,
Pakizeh2008, Ponsinet2015} particularly those containing metals that have been
structured in some way. The mechanism is familiar from the concept of
metamaterials whereby novel electrical and magnetic properties can be found by
engineering the microstructure
\cite{Pendry1996,Pendry1999, Smith2000, Merlin2009}. Here
we present an experimental study of a structured system, in our case a grating,
and interpret the data in terms of effective parameters of an equivalent
ultrathin layer of metamaterial. Features such as asymmetric reflection will emerge
from the interplay of magnetic and electric fields.\\

Lately, a lot of attention has been given to the development of metasurfaces --
textured surfaces at a sub-wavelength scale with tailored electro-magnetic
properties \cite{Holloway2012, Kildishev2013, Shi2014, Meinzer2014, Stauber2014,
Genevet2015}. These devices are the two-dimensional counterpart of
metamaterials and benefit from the advantages of reduced profile, reduced ohmic and magnetic losses, and
simpler fabrication processes, thus enabling a range of applications from analogue computing \cite{Silva2014}, to `flat optics' \cite{Yu2014} and light harvesting applications \cite{Atwater2010, Wu2012}, to name a few. \\

A special class of metasurfaces comprises bianisotropic ones, that is, materials
exhibiting coupling between the electric and magnetic fields \cite{Kongbook, Kildishev2011, Eliane2010,
Medina2002}, which are described by the constitutive relations
\cite{Kongbook}:
$\mathbf{D} =\boldsymbol{\epsilon}\mathbf{E}+\boldsymbol{\xi}\mathbf{H}$ and
$\mathbf{B} =\boldsymbol{\mu}\mathbf{H}+\boldsymbol{\zeta}\mathbf{E}$ (see
methods), with the coupling parameters $\boldsymbol{\xi}$  and $\boldsymbol{\zeta}$
relating the displacement current (D) with the magnetic field (H), and the
magnetic flux density (B) with the electrical field (E), respectively.
A typical  characteristic of bianisotropic structures is their asymmetric
response depending on which side of the structure they are being probed from. 
\cite{LiBiAn}

There is a large body of literature, discussing bianisotropy in the THz \cite{Eliane2010,
Radi2015} and GHz regimes\cite{Yazdi2015, Aydin2007, Pfeiffer2014, Zhao2015}, with applications in polarization changers
\cite{Pfeiffer2014,Zhao2012}, asymmetric absorbers/reflectors \cite{Yazdi2015,
Zhao2015} and perfect absorbers \cite{Radi2013}.
However,  in the spectral optical regime, very few experimental studies exists \cite{Zhao2012,Pfeiffer2014a,Butun2015}. 
Recently, an asymmetric absorber operating in the visible frequency range has
been designed \cite{Butun2015}, where the asymmetry originates from a
dielectric-metal stack of thin a Ag nano-holes array patterned over a Silicon-Nitride membrane.\\ 
In this paper we  show experimentally that asymmetric reflection
based on bianisotropy and magneto-electric coupling can be achieved at visible
wavelengths simply, using all-metal plasmonic gratings with reduced thickness of the
active region, in a design that requires simple and commonly used fabrication methods. Electrodynamic simulations are
compared with experimental results and are interpreted in terms of the
effective material parameters of the system, which clearly indicate magneto-electric coupling.

\section{Results and discussion}

\begin{figure}[t!]
    \centering
    \includegraphics[width=0.4\textwidth]{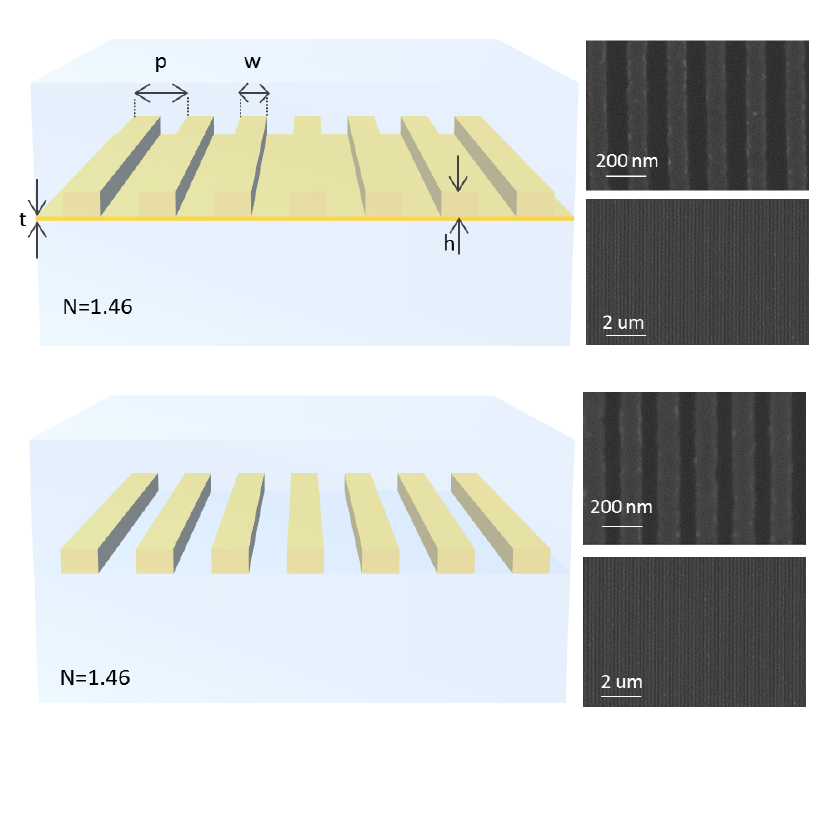}
    \caption{The grating structures. Top: the bianisotropic structure.  The letters and 
             arrows indicate the physical dimensions of the structure: w=100nm is
             the width of grating lines; p=200nm is the grating pitch; t=10 nm is
             the thickness of the gold layer; and h=55nm is the thickness of the
             grating lines (including the 10nm gold layer which covers both the grating lines and the spacing between them). Bottom: the 
             control structure with w=110nm; p=200nm, and h=45nm.  Both structures are embedded
             within a medium with n=1.46 at $\lambda$=560nm. Placed to the 
             right of the illustrations are zoomed-out and zoomed-in SEM images 
             of the grating structures, taken before the deposition of the thin-film 
             of gold and the attachment of the covering quartz-slide.}
    \label{fig:Samples}
\end{figure}

Our bianisotropic metamaterial structure is made of metallic grating lines and a
thin conducting layer covering the grating lines and the spacing between them. 
Figure \ref{fig:Samples} illustrates the bianisotropic structure
-- 80$\mu$m$\times$80$\mu$m arrays of a 50nm thick gold grating covered by 10nm
layer of gold -- next to  a `control' structure with a similar grating array, but lacking the thin gold
layer. The two types of gold structures were fabricated by means of electron beam
lithography and thermal evaporation on the same quartz substrate. Finally, in order
to limit asymmetries in the device only to the active metallic gratings
(and not to the substrate, for example) the device was completed by the bonding of
another quartz slab on top of the structures (see Methods for fabrication details). \\

\subsection{Reflection and Transmission spectra}
\begin{figure}[t!]
    \centering
    \includegraphics[width=0.4\textwidth]{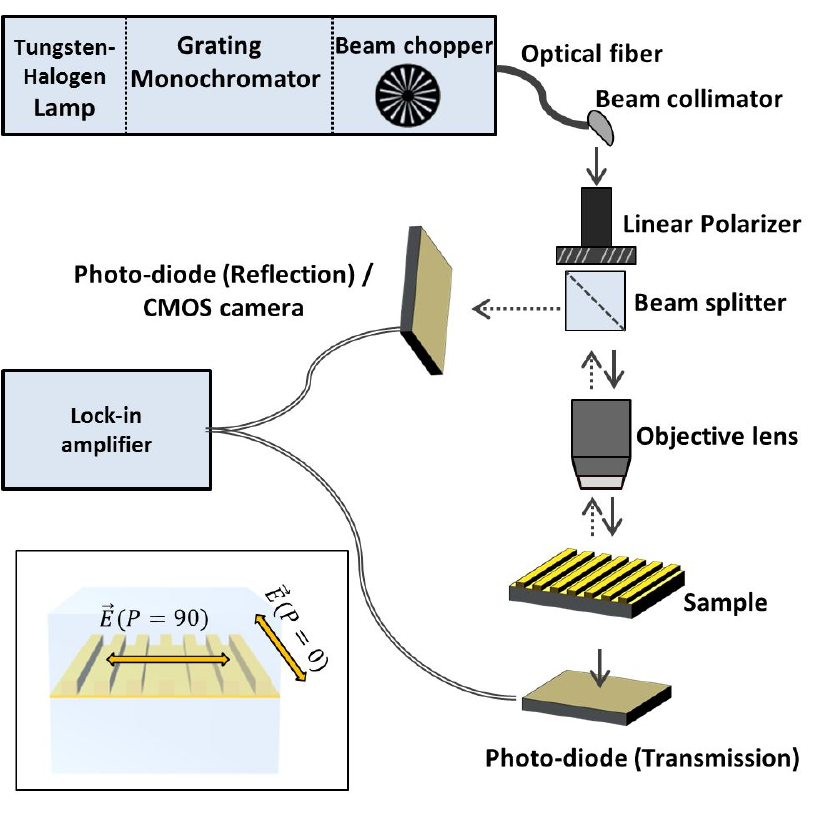}
    \caption{Experimental set-up for transmission and reflection measurements.
             White light is being diffracted, chopped and coupled to an optical
             microscope via an optical fiber and beam collimator. The collimated monochromatic
             light was linearly polarized (along or transversely to the gratings
             lines) and focused on the structures. A beam-splitter was used to 
             direct the reflected light toward the reflection photo-diode (the 
             path of the reflected light is indicated by dashed arrows), or toward
             a camera used to assist in locating the structures on the sample. The transmitted/reflected light was recorded on Si photo-diodes 
             connected to a Lock-in amplifier.}
    \label{fig:Exp_setup}
\end{figure}

\begin{figure}[t!]    
  \centering
  \includegraphics[width=1\textwidth]{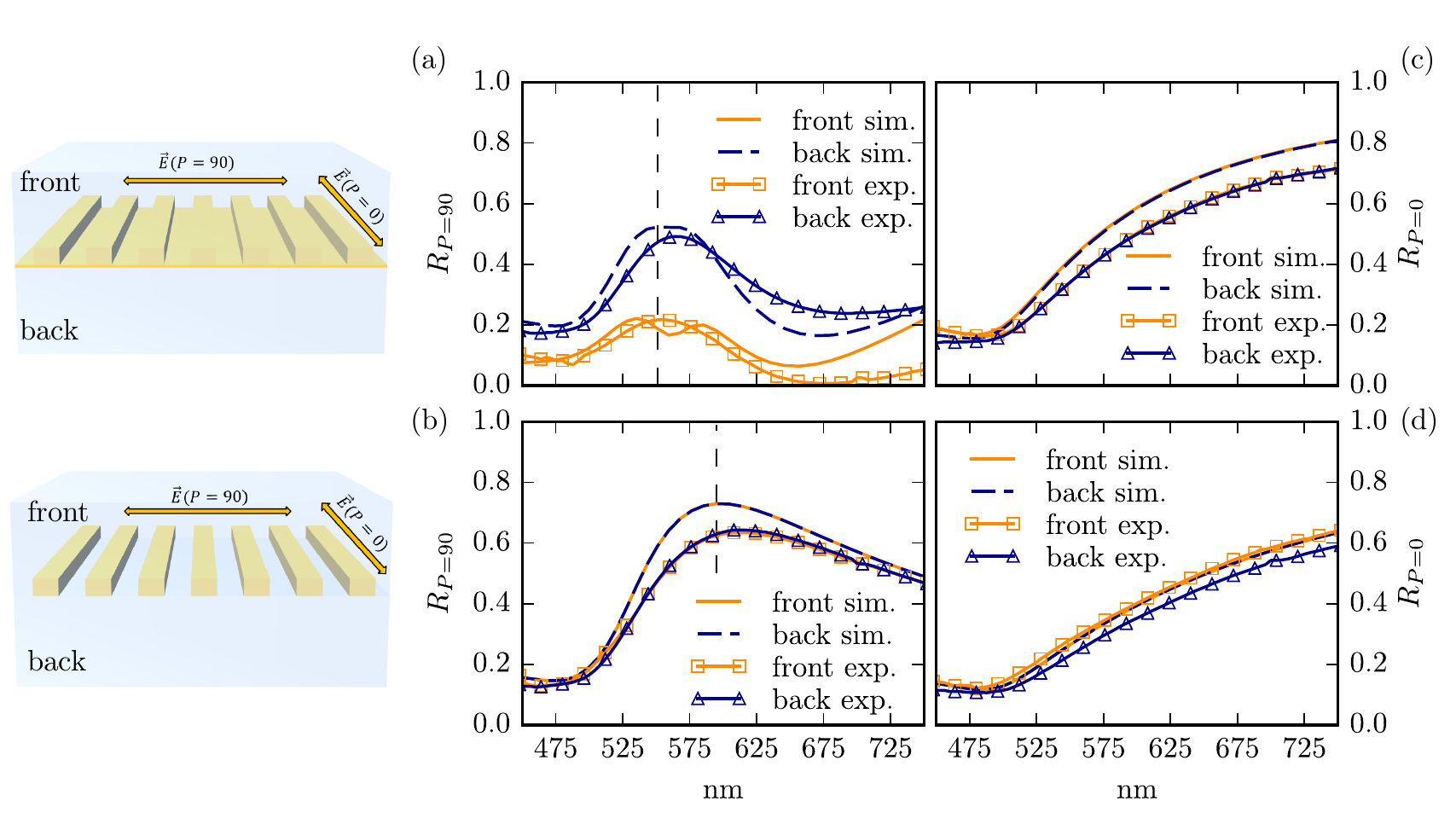}
  \caption{
          Reflectance spectra. Open triangles and squares are the experimental
          results for `back' and `front' exposure, as defined in the
          illustration on the left to the plots; blue dashed and orange solid
          lines correspond to numerical simulations of `back' and `front'
          exposures. a) bianisotropic structure probed with orthogonal
          polarization of the electric field (P=90). c)
          bianisotropic structure excited with polarization of the electric
          field along the grating lines (P=0). b) and d) the control structure with metal grating only
          for electrical-field polarizations  P=90, and P=0, respectively. The
          vertical dashed lines indicates the spectral positions of the resonant
          peaks.}
  \label{fig:R_GAuG}
\end{figure}

The reflection and transmission spectra of the two types of structures were 
obtained using a setup of an optical microscope coupled to a grating 
spectrometer for two linear polarizations: one for which the electric field is 
aligned with the grating lines ($P=0$) and one where the polarization is 
orthogonal to the gratings lines ($P=90$).
For both simulations and measurements the plane 
wave is normally incident (note that in our experiments, though centered around 
the normal to the plane of the gratings, the plane wave is distributed throughout the
numerical aperture of the objective (N=0.5)). The recording of the transmitted 
and reflected light was completed via Si photo-diodes coupled to a lock-in amplifier
(Figure \ref{fig:Exp_setup}; see Methods for detailed information about the 
measurements set-up).\\

\begin{figure}[t!]
    \centering
    \includegraphics[width=0.9\textwidth]{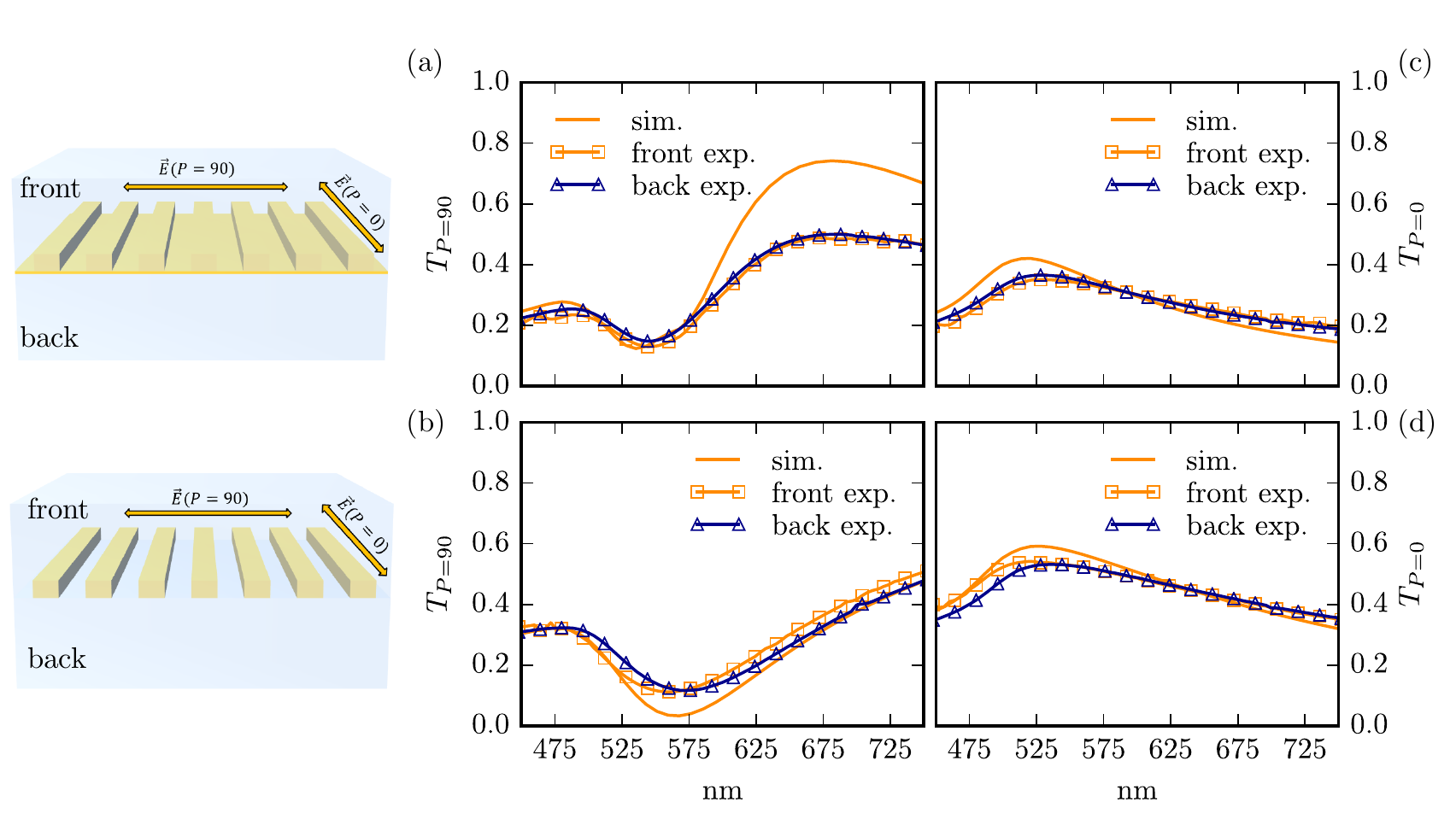}
    \caption{
            Transmission spectra. Open triangles and squares are the
            experimental results for `back' and `front' exposure, as defined in
            the illustration on the left to the plots; blue dashed and orange
            solid lines correspond to numerical simulations of `back' and
            `front' exposures. a)  bianisotropic structure probed with
            orthogonal polarization of the electric field (P=90 in
            Figure \ref{fig:Exp_setup}). c) bianisotropic structure excited with polarization of
            the electric field along the grating lines (P=0 in
          Figure   \ref{fig:Exp_setup}). b) and
            d) reflection spectra of the control structure with metal grating only for electrical field
            polarizations  P=90, and P=0, respectively.
            }
    \label{fig:T_GAuG}
\end{figure}

Figure \ref{fig:R_GAuG} shows the experimental and simulated reflectance spectra
for the bianisotropic and the control structures. For polarized light with its
electric field \emph{orthogonal} to the grating lines it shows a remarkable
difference between the control sample with the metal gratings only, and the
bianisotropic sample with the additional thin layer of gold: while the
reflection spectra of the control structure (Figure \ref{fig:R_GAuG}c) are inherently
indifferent to the direction of incoming light, as expected from an optically
symmetric structure, the reflection signal of the gold grating with the
additional thin gold layer (Figure \ref{fig:R_GAuG}a ) is up to $\approx$3 times
stronger when the device is probed from the side of the thin-film (aka, `back
exposure'), compared to probing from the corrugated side ('front exposure'). For
this polarization of the electrical field the bianisotropic and the control
devices show a plasmon resonance with maximal reflection at $\approx$550nm and
$\approx$595nm, respectively -- indicating strong coupling to the incident
light. When probing the metal grating with the electrical field polarized
\emph{along} the grating lines,  (Figure \ref{fig:R_GAuG}b and d), both the plasmon
resonance and the asymmetric reflection diminishes, and we are left with a
reflection curve akin to that of a planar Au layer.  Unlike the reflection spectra, the transmission curve
(Figure \ref{fig:T_GAuG}) is independent of the illumination
direction, regardless to structure and polarization, as it must be due to
reciprocity. For all structures and polarizations we see a qualitative
agreement between experimental and numerical results with quantitative
differences arising most likely from fabrication imperfections and uncertainties
in the value of the permittivity.

It appears that an asymmetric reflection from our low profile structures is
evident when the following two conditions are satisfied simultaneously: a) the
device has to bear some optical asymmetry, and b) the incoming light has to be
strongly coupled to, and absorbed by the active part of the device -- here,
achieved by means of plasmonic coupling between the incident light and the
grating, when probed with the appropriate polarization. 
It is worth noting that, in \emph{bulk} materials and structures, an optical asymmetry -- and therefore asymmetric
reflection -- is abundant in nature and our daily life. Take for example a single
side polished Silicon wafer, or a common bathroom mirror; both meet the former
condition above and demonstrate asymmetric reflection with one side reflecting
more than the other. However, while the asymmetric reflection in the case of the two macroscopic examples given above is due to  their two dissimilar non-interacting surfaces, achieving asymmetric reflection while maintaining
a low structure profile at the order of $\lambda$/10, as demonstrated here, requires a design which meets also the second
condition of strong coupling between the incoming light and the optically interacting surfaces.

As asymmetric reflection could only be observed for the grating with the 10nm
connecting layer of gold, it is clear that understanding the role of the thin
conducting layer is crucial for the understanding of the origin of the
asymmetric reflection: the conductive layer electrically connects neighboring
grating lines, therefor enabling a flow of electrical current in the direction
normal to the grating lines (P=90). The induced current, in turn,
generates a magnetic field \cite{Kraft2015} radiating energy into the far field
and interfering with the incident wave. This effect is enhanced via a plasmon
resonance, as more energy is coupled into the grating. However, the structural
asymmetry leads to different coupling strengths for front or back illumination.
If the light is incident on the flat side, the surface does not immediately
provide the missing momentum to couple to plasmons so that a large fraction of
the light is reflected. On the corrugated side, the surface immediately
provides the missing momentum and coupling into plasmons is more efficient
leading to higher absorption and thus reduced reflection properties. This behavior is
well known from bianisotropic bulk materials exhibiting magneto-electric
coupling \cite{Kongbook, LiBiAn}.
In order to further investigate the magneto-electric coupling and quantify its
strength, we continue by modeling our meta-surface as a bianisotropic
metamaterial \cite{Kongbook, LiBiAn}.

\subsection{Effective material parameters and bianisotropy}
\begin{figure}[t!]
   \centering
   \includegraphics[width=1\textwidth]{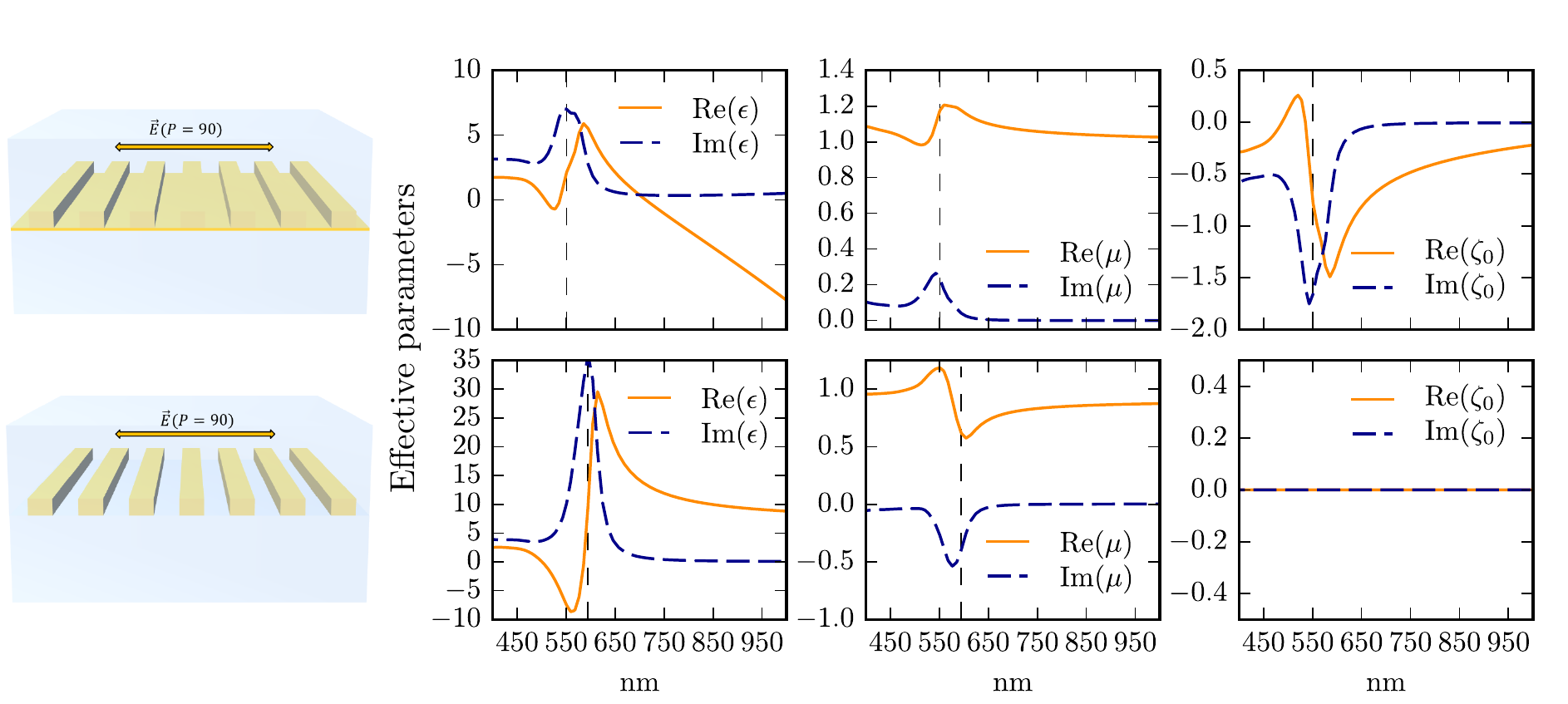}
   \caption{
            Effective material parameters retrieved from numerical simulations
            of the gold gratings described in figure \ref{fig:Samples}. Results
            are for $P=90$. The left column shows the effective permittivity,
            the middle column the effective permeability and the right
            column the bianisotropic coupling parameter. We used standard retrieval methods \cite{Chen2005,
            LiBiAn,Eliane2010} and assumed a symmetric metamaterial slab with
            thickness $d_{eff}$ equal to the maximum thickness of the gold
            gratings, i.e. $d_{eff}=55$nm for the connected (top) and
            $d_{eff}=45nm$ for the unconnected grating (bottom). The vertical
            dashed lines indicate the positions of the respective reflection
            maxima. 
           }
   \label{fig:Param_P90}
\end{figure}

An asymmetric reflection/absorption spectrum cannot be described by a
\emph{homogeneous} slab model using effective permittivity and permeability
alone;  instead, magneto-electric coupling has to be introduced to the physical
model by modifying the constitutive relations \cite{Kongbook, LiBiAn,
Eliane2010}. The magneto-electric coupling is quantified by new material
parameters, $\boldsymbol{\zeta}$ and $\boldsymbol{\xi}$ (see introduction and
section \ref{sc:methods_numerical}), which make the electric and magnetic fields
interact in a way such that even a symmetric slab exhibits asymmetric
reflection \cite{Kongbook, LiBiAn, Eliane2010}.\\

We model our connected grating as a homogeneous bianisotropic slab and
compare against the control grating. Note that the magneto-electric coupling
parameters $\boldsymbol{\xi}$ and $\boldsymbol{\zeta}$ are determined by a
single parameter $\zeta_0$, on which we focus hereafter.
Figure \ref{fig:Param_P90} shows the material parameters retrieved from
simulations of the types of gold gratings structures for $P=90$. Clearly, the interplay of
plasmon resonance and structural asymmetry in the connected gold grating lead to
an induced magneto-electric effect, as demonstrated by the strong resonance in the coupling parameter 
$\zeta_0$. The control grating, however, shows no such coupling, highlighting
the importance of that thin connecting Au layer for the design. 
Figure \ref{fig:Param_P90} exhibits two additional differences between the bianisotropic and control
gratings. First, though both samples feature optical magnetic effects (resonant
permeability) at the plasmon resonance, the permeability of the control grating
does not return to unity in the long wavelength limit, but remains weakly
diamagnetic. This seemingly counter-intuitive result is explained as follows. In
the long wavelength limit, gold approaches a perfect conductor, thus
expelling part of the magnetic field from its interior. This results in
`squeezing' of the magnetic field lines through the gaps of the grating, leading
to the diamagnetic effect observed \cite{Wood2007}.The smaller the gap,
the stronger this effect (see figure S3 in the supplementary material).

The case of the bi-anisotropic sample with the connected grating is
understood by analogy with a superconductor where the permeability is
unity, but the permittivity diverges inversely as the frequency squared.
In this case magnetic fields are expelled entirely through the action of
the permittivity, contrary to some misinformed statements to the effect
that superconductors are perfect diamagnets. In figure 5 top left we see
that for the continuously connected sample the effective permittivity
diverges at long wavelengths.

\begin{figure}[t!]
   \centering
   \includegraphics[width=\textwidth]{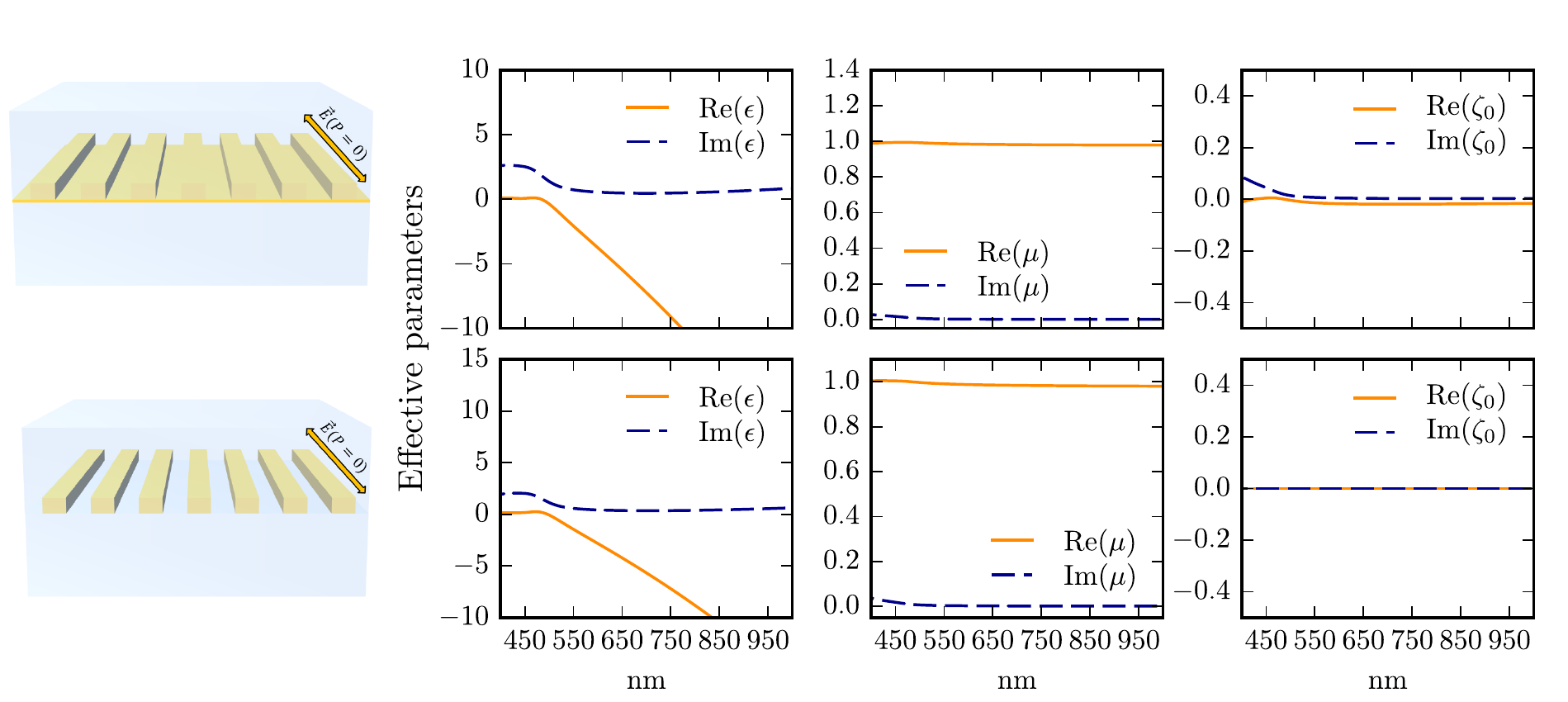}
   \caption{
            As figure \ref{fig:Param_P90} but for $P=0$.
           }
   \label{fig:Param_P0}
\end{figure}

In figure \ref{fig:Param_P0} we look at the effective material parameters for
$P=0$.  
Here, the electric field is polarized along the grating lines and current can flow freely in both the control and bi-anisotropic 
gratings, similar to the current flow in thin gold films.
The lack of a plasmon resonance leads to a near-zero magneto-electric coupling for the connected gold grating and the reflection spectra only shows minor deviations at short wavelengths at the order of the grating period.

\section{Conclusion}
We demonstrated experimentally and through simulations that
plasmonic gratings provide a simple design for asymmetric reflectors/absorbers
at visible wavelengths. We found that the asymmetry in the spectra is a direct
consequence of magneto-electric coupling in the grating, i.e. bianisotropy, and
should be interpreted as such. The strength of the bianisotropy provides a good
measure for the expected asymmetry in the spectra and may be used for the
optimization of this effect, as has been done at much lower frequencies
previously \cite{Yazdi2015}. Finally, the bianisotropy of our grating is inherent to it,
rather than induced by a substrate \cite{Butun2015}, which leads to an extremely
small footprint of the structure ($\approx 55nm$ thickness). Further
optimizations in the grating design and fabrication are certain to enhance the
strength of the asymmetry.

\section{Methods}

\subsection{Experimental setup and fabrication}
\textbf{Fabrication of gratings structures:}
First, quartz substrates were cleaned by ultrasonic agitation in an acetone bath, followed up by IPA and DIW rinse.
A Polymethyl methacrylate (PMMA 950, A4) positive resist spun on the cleaned substrates at 3500RPM for 60 seconds, and soft-baked on a hotplate at 180 C for 60 seconds, forming a ~250 nm lift-off mask. 
Multiple arrays of grating lines were defined using a Raith eline Electron Beam Lithography tool at 10 kV accelerating voltage and 10 micro-meter aperture, and developed for 30 seconds in 1:3 methyl isobutyl ketone:isopropyl alcohol (MIBK:IPA) developer. 
Prior to the deposition of gold grating lines, the surface was modified to improve the adhesion of gold to the SiO2 substrate: a drop of (3-aminopropyl)triethoxysilane diluted in ethanol at a volumetric ratio of 1:25 was dropped and was left on the sample for 30 seconds before washed in ethanol and dried on a hotplate at 90C for 15 minutes.
45 nm of Au were thermally evaporated (Angstrom AMOD) at a rate of 2A/s. The fabrication of the grating pattern was completed by lift-off of the gold layer in acetone bath for 24H at room-temperature. 
After lift-off, the surface was functionalized again (as discussed above), and an additional layer of 10nm of Au was deposited on the sample (at a higher deposition rate of 5A/s to improve surface uniformity) through a mechanical mask that shaded half of the sample, completing the control, and bianisotropic structures on a single substrate. 
The device was finished by bonding an additional quartz substrate on top, using a UV cured, indexed matched, low viscosity optical glue (Norland Optical Adhesive 85). 

\textbf{Reflection transmission measurements} 
were carried out using a custom-made microscope set-up: The output of  a Halogen lamp connected to a Bentham Instruments monochromator was coupled to a multimode optical fibre with a reflective collimator at its end. The collimated beam passed through a  Glan-Thompson polarizer (Thorlabs) and a non-polarizing beamsplitter cube (coated with anti reflective-coating optimized to 400-700nm; Thorlabs) before entering a custom built microscope system and focused to a spot of diameter ~50$\mu$m via 50X long-working distance objective lens with NA=0.5 (see Figure \ref{fig:Exp_setup}). 
The reflected and transmitted spectra were recorded using a Si photo-diode connected to a lock-in amplifier. Absolute transmission values were calculated by dividing the measured transmission photo-current by the measured current without the sample.  Absolute reflection values were calculated by first subtracting  background reflections (originated mainly from the beam-splitter for wave-lengths below $\sim$450nm or above $\sim$680nm; measured for the same optical path without the sample) from the reflection photo-current, and from the reflectance measured from a calibrated silver coated mirror, and then dividing the back-ground-free reflectance current of the sample by that of the calibrated mirror.

\subsection{Numerical modeling}\label{sc:methods_numerical}
The numerical simulations were carried out using the frequency domain solver of
the commercial software package COMSOL Multiphysics (version 4.3a). We used the
port boundary conditions to set up the exciting plane wave and periodic boundary
conditions for the other two boundaries. The length of the simulation domain has
been set to $1500$nm and convergence with respect to it had been ensured. We
used an adaptive triangular mesh with maximum element size of $2$nm inside the
gold gratings and $10$nm in the surrounding medium; again convergence had been
ensured. The exact dimensions of the gratings were chosen in accordance with the
experimental structures; detailed sketches can be found in the supplementary. 
We used experimental values for the permittivity of an evaporated gold film from
\cite{Olmon}.
The port boundary conditions allow to compute the scattering, i.e. S-parameters
of the system, from which reflectance and transmittance can be deduced, as well
as the effective material parameters
\cite{LiBiAn,Chen2005, Eliane2010}.\\
In this letter we modeled the two gratings as a bianisotropic metamaterial. That
is, we replaced the whole system with a symmetric metamaterial slab of thickness
$d_{eff}$ and effective material parameters determined by the scattering
parameters of the gratings. Bianisotropy means that the electric and magnetic
fields inside such a material are coupled via \cite{Kongbook, LiBiAn}

\begin{align*}
    \mathbf{D} & =\boldsymbol{\epsilon}\mathbf{E}+\boldsymbol{\xi}\mathbf{H}\\
    \mathbf{B} & =\boldsymbol{\mu}\mathbf{H}+\boldsymbol{\zeta}\mathbf{E},
\end{align*}

\noindent with permittivity and permeability tensor \cite{Kongbook, LiBiAn}

\begin{equation}
    \boldsymbol{\epsilon}=
        \left(\begin{array}{ccc}
             \epsilon_{||} & 0 & 0\\
             0 & \epsilon_{||} & 0\\
             0 & 0 & \epsilon_{\perp}
        \end{array}\right),
    \boldsymbol{\mu}=
        \left(\begin{array}{ccc}
            \mu_{||} & 0 & 0\\
            0 & \mu_{||} & 0\\
            0 & 0 & \mu_{\perp}
        \end{array}\right)
\end{equation}

\noindent and magneto-electric coupling matrices \cite{Kongbook, LiBiAn}
,

\begin{equation}
    \boldsymbol{\xi}=
        \left(\begin{array}{ccc}
            0 & 0 & 0\\
            0 & 0 & 0\\
            0 & -i\xi_{0} & 0
        \end{array}\right),      
    \boldsymbol{\zeta}=
        \left(\begin{array}{ccc}
            0 & 0 & 0\\
            0 & 0 & i\xi_{0}\\
            0 & 0 & 0
        \end{array}\right).
\end{equation}

\noindent where $||$ and $\perp$ stands for in- and out-of-plane respectively.
The magneto-electric coupling parameter that has been retrieved in this letter
corresponds to the effective $\zeta_0$. The effective permittivity corresponds
to the in-plane component of the permittivity tensor and the effective
permeability to the out-of-plane component of the permeability tensor.

Standard methods have been developed to retrieve these parameters for a
symmetric metamaterial slab from the S-parameters obtained from COMSOL
\cite{LiBiAn,Chen2005, Eliane2010}.
In our case, the thickness, $d_{eff}$, of this metamaterial slab was chosen to
be the maximum thickness of the gold gratings.

\begin{acknowledgement}
The authors gratefully acknowledge support from the EPSRC EP/L024926/1
programme grant and the Leverhulme Trust. M.K. acknowledges support 
from the Imperial College PhD scholarship. A. B. acknowledges support
from the Imperial College London Junior research fellowship.
Y.L. acknowledges NTU-A*start Silicon Technologies of Excellence under
the program grant No. 11235150003.
S.A.M acknowledges the Royal Society and the Lee-Lucas Chair in Physics.
J.B.P thanks the Gordon and Betty Moore foundation.
\end{acknowledgement}

\begin{suppinfo}
    A supplementary file is provided with this manuscript. The file
    contains detailed profiles of the gratings used in the simulations,
    a figure showing the mode profile at the plasmon resonance and
    a figure comparing the retrieved effective permeability for 
    control gratings with different gap sizes between the grating lines.
\end{suppinfo}
\bibliography{Main_bibliography}


\end{document}